\documentclass[twocolumn,showpacs,preprintnumbers,amsmath,amssymb]{revtex4}
\usepackage{epsfig}
\usepackage{dcolumn}
\usepackage{bm}
\newcommand{\HR}[1]{{#1}}

\begin{document}

\title{Self-sustained activity in a small-world network of excitable neurons}
\author{Alex Roxin$^+$, Hermann Riecke$^+$ and Sara A.~Solla$^*$}
\affiliation{$^+$Engineering Science and Applied Mathematics, 
$^*$Department of Physiology and Department of Physics and Astronomy, Northwestern University, Evanston, IL 60208, USA}
\date{\today}

\begin{abstract} 

We study the dynamics of excitable integrate-and-fire neurons in
a small-world  network.  At low densities $p$ of directed random
connections, a localized  transient stimulus results in either
self-sustained persistent activity or in a brief transient
followed  by failure. Averages over the quenched ensemble reveal
that the  probability of failure changes from $0$ to $1$ over a
narrow range in $p$;   this failure transition can be described
analytically through an extension of  an existing mean-field
result. Exceedingly long transients emerge at higher densities $p$;
their activity patterns are disordered, in contrast
to the mostly periodic persistent patterns observed at low
$p$.   The times at which such patterns die out \HR{are
consistent with}  a  stretched-exponential distribution, which
depends sensitively on the propagation velocity of the excitation.  

\end{abstract}  
\maketitle

Recent research in complex networks has provided increasing 
evidence for their relevance to a variety of physical, biological, 
and social phenomena \cite{WaSt98,St01,AlBa02}. Two distinct types 
of topology have been particularly useful in providing insights into
the implications of complex connectivity:  scale-free networks
\cite{AlBa02}, characterized by the existence of a small number
of hubs with high coordination number, and small-world networks
\cite{WaSt98}, characterized by the presence of shortcuts that 
link two randomly chosen sites regardless of the distance between 
them. 

So far, most work on complex networks has focused on their 
topological and geometrical properties; less attention has been
given to the properties of dynamical systems defined on such
networks. The interplay between the intrinsic dynamics of  the
constituent elements and their complex pattern of connectivity
strongly affects the collective dynamics of the resulting
system. For instance, the addition of shortcuts induces a 
finite-temperature phase transition even in the one-dimensional
Ising model \cite{Pe01} and the introduction of unidirectional
shortcuts can change the second-order phase transition in the
two-dimensional Ising model into a first-order one
\cite{SaLo02}.  In a system of coupled oscillatory elements, the
introduction of shortcuts enhances  synchronization 
\cite{BaPe02},  while the introduction of hubs eliminates the
threshold for epidemic propagation \cite{PaVe01}. 

The coexistence of shortcuts and regular local connections
characteristic of small-world networks (SWNs) mimics a salient
feature of the circuitry in the cortex
\cite{Mo97,GoBa00,McCo01,WuGu99,DeHa01,MeGo01},  where
experimental observations of excitatory traveling waves
\cite{WuGu99}  provide evidence of some degree of local
connectivity, while it is also recognized that long-range
excitatory connections are present \cite{DeHa01,McCo01}. Our
goal is to explore the influence of this complex connectivity on
the dynamics of neuronal circuits; to this purpose, we choose a
minimal model. The underlying network is modeled as a SWN with
unidirectional shortcuts that reflect the  nonreciprocal
character of synaptic connections, and the excitable  neurons
are modeled as leaky integrate-and-fire units. We find that 
even this simple model exhibits a rich repertoire of distinct
dynamical behaviors as a function of the density $p$ of added
shortcuts: a low $p$ regime characterized by persistent periodic
activity that is bistable with the quiescent state, a transition
to failure with increasing $p$, followed by a reemergence of
long-lasting disordered activity. We note that a SWN with
unidirectional shortcuts has been considered in a different
regime by  Lago-Fern\'{a}ndez et al \cite{LaHu00,LaCo01} to
address the possibility of rapid synchronization among
conductance-based neurons of the Hodgkin-Huxley type.

The model considered here consists of a one-dimensional array of $N$
integrate-and-fire neurons (IFNs) in which a SWN topology is
created through the addition of a density $p$ of unidirectional 
long-range couplings.  The membrane potential of the IFNs is
determined by  
\begin{eqnarray} 
\tau_{m}\frac{dV_{i}}{dt} =
-V_{i}+I_{ext}+
g_{syn}\sum_{j,m}w_{ij}\delta
(t-t_{j}^{(m)}-\tau_{D}).  \label{Veq}  
\end{eqnarray}  
A neuron fires whenever its voltage exceeds $1$; the voltage is
then  reset to $0$. The external current is chosen to satisfy
$I_{ext} < 1$; in this regime the IFNs are not oscillatory, but
excitable. The last term models input currents due to
presynaptic firing as  a delayed  impulse: if $w_{ij}$ is
non-zero, then neuron $i$ receives a pulse input of amplitude
$g_{syn}$ with a delay $\tau_{D}$ after neuron $j$ has fired its
$m$th  spike at time $t_{j}^{(m)}$.  The synaptic conductance is
chosen to  satisfy $I_{ext}+g_{syn} > 1$, so that a single
input  suffices to sustain firing activity. The local
connections are modeled here as nearest-neighbor  couplings
($w_{i,i\pm 1}=1$) that define an underlying regular lattice. The
long-range connections result from randomly {\it adding} rather 
than rerouting \cite{WaSt98} a fixed fraction $pN$ of unidirectional 
couplings $w_{ij}=1$  to generate a SWN topology.

\begin{figure}
\centerline{
\epsfxsize=3.0in{\epsfbox{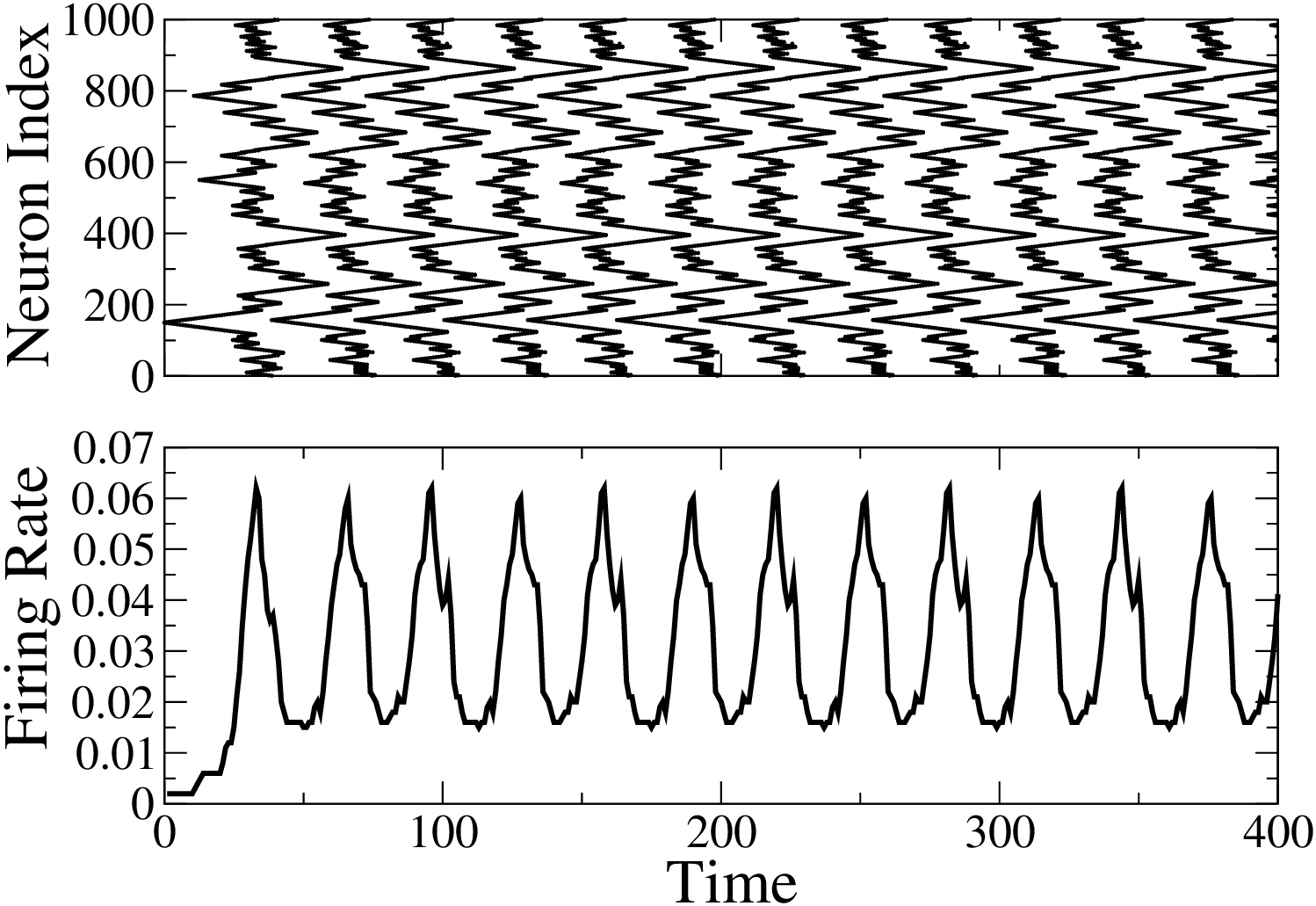}}}
\caption{Raster plot (top) and instantaneous firing rate (bottom) 
for a system with $N=1000$, 
$I_{ext}=0.85$, $g_{syn}=0.2$, $\tau_{m}=10$,
$\tau_{D}=1$, $p=0.1$. Same parameter values are used in subsequent figures 
unless noted otherwise.} \label{fig1}
\end{figure}

\begin{figure}
\centerline{
\epsfxsize=3.0in{\epsfbox{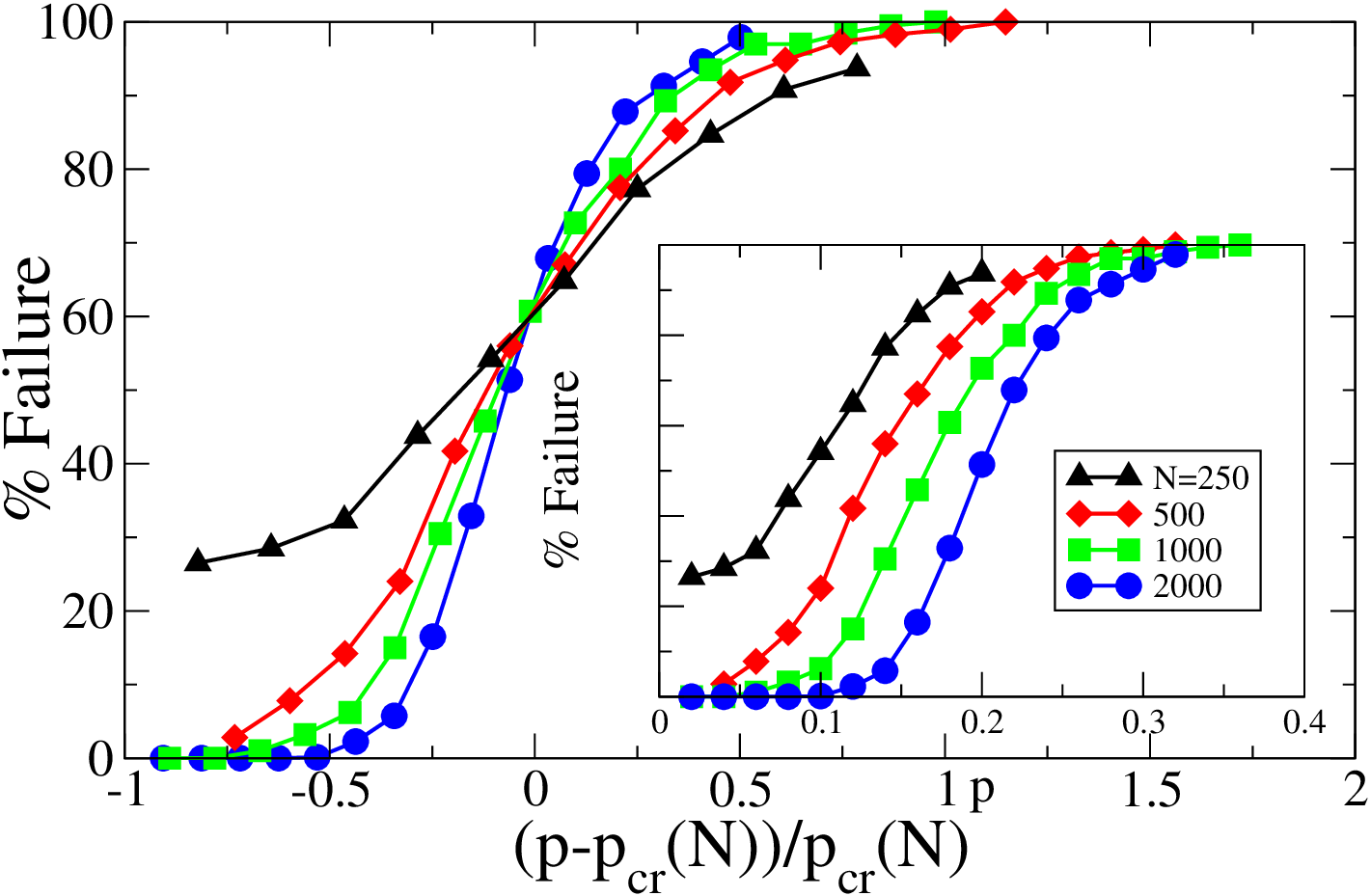}}}
\caption{Inset: Failure rates for $\tau_D=1.0$ and different system sizes $N$. 
Main: Failure rates normalized by $p_{cr}(N)$
(cf.(\ref{crit},\ref{Ta})).} \label{fig2}
\end{figure}

At $p=0$, any excitation sufficient to cause a neuron to fire
will generate two pulses that propagate through the regular
lattice  in opposite directions with velocity $v=1/\tau_{D}$,
and  either exit the system or annihilate each other, depending
on boundary  conditions. No persistent activity results in
either case. However, self-sustained activity may arise for
nonzero $p$, as shown in  fig.\ref{fig1}. Persistence relies on the
reinjection of activity via a shortcut into a previously
active domain that has by then recovered; this reinjection can occur 
only if the shortcuts are unidirectional. For a fixed value of $p$, 
any particular network realization has a different connectivity graph 
that may or may not sustain persistent activity. We average over typically
2000 realizations to calculate the probability of persistent
activity; the complementary probability  of \textit{failure} to
sustain activity is shown in fig.\ref{fig2}  (inset) as a
function of the density $p$ of random  connections for four
different system sizes.  In this regime, the probability of
failure is an increasing function of $p$ that crosses over from
$0$ to $1$ with increasing steepness as the size $N$ of the
system increases. 

Failure to sustain activity is a simple consequence of the
intrinsic dynamics of the neurons. Pulses travel outwards from
an initial activity seed, and spawn new pulses at a rate that
increases with $p$. A currently inactive neuron can fire
again only if enough time has elapsed  from its preceding
firing to allow for a recovery to  $V\ge 1-g_{syn}$.  A single
input will be able to elicit a spike only if the elapsed
time exceeds $T_R^{(1)}$, with  
\begin{equation} 
T_{R}^{(n)} =
\tau_{m}\ln{\Big( \frac{I_{ext}}{I_{ext}+n g_{syn}-1}\Big)}.
\label{Tr} 
\end{equation} 
If activity recurs to a given site too
rapidly, the neuron will fail to produce a spike, and the
pulse of activity will die out.  A critical density $p_{cr}$ for the
transition from self-sustained activity to failure can be
estimated from 
\begin{equation} 
T_{A}(p_{cr}) = T_{R}^{(1)}
\label{crit},  
\end{equation} 
where $T_{A}(p)$ is the time
needed for the activity to spread across the whole network. At
a fixed velocity for pulse propagation, this time corresponds
to the largest distance across the network. This distance has
been calculated for bidirectional shortcuts using a mean-field
approach \cite{NeMo00}; when extended to the case of
unidirectional shortcuts it results in  
\begin{equation}
\sqrt{\Big(1+\frac{4}{pN}\Big)}\tanh{\Big[\sqrt{\Big(1+\frac{4}{pN}\Big)}\frac{pT_{A}(p)}
{2\tau_{D}}\Big]}=1.  
\label{Ta} 
\end{equation} 
An implicit expression for $p_{cr}$ and its dependence on the
system size $N$ and the propagation velocity $v$ follows from
combining equations (\ref{crit}) and (\ref{Ta}); for large $N$, 
$p_{cr}(N)\propto \ln{N}$. Failure rate curves as a function of
$(p-p_{cr}(N))/p_{cr}(N)$, shown in fig.\ref{fig2}, cross at
the  theoretically predicted value. This observation, together
with  the increased steepness of these  curves with increasing
$N$, indicates that a well-defined transition to failure occurs
in the thermodynamic limit. 

\begin{figure}
\centerline{
\epsfxsize=3.0in{\epsfbox{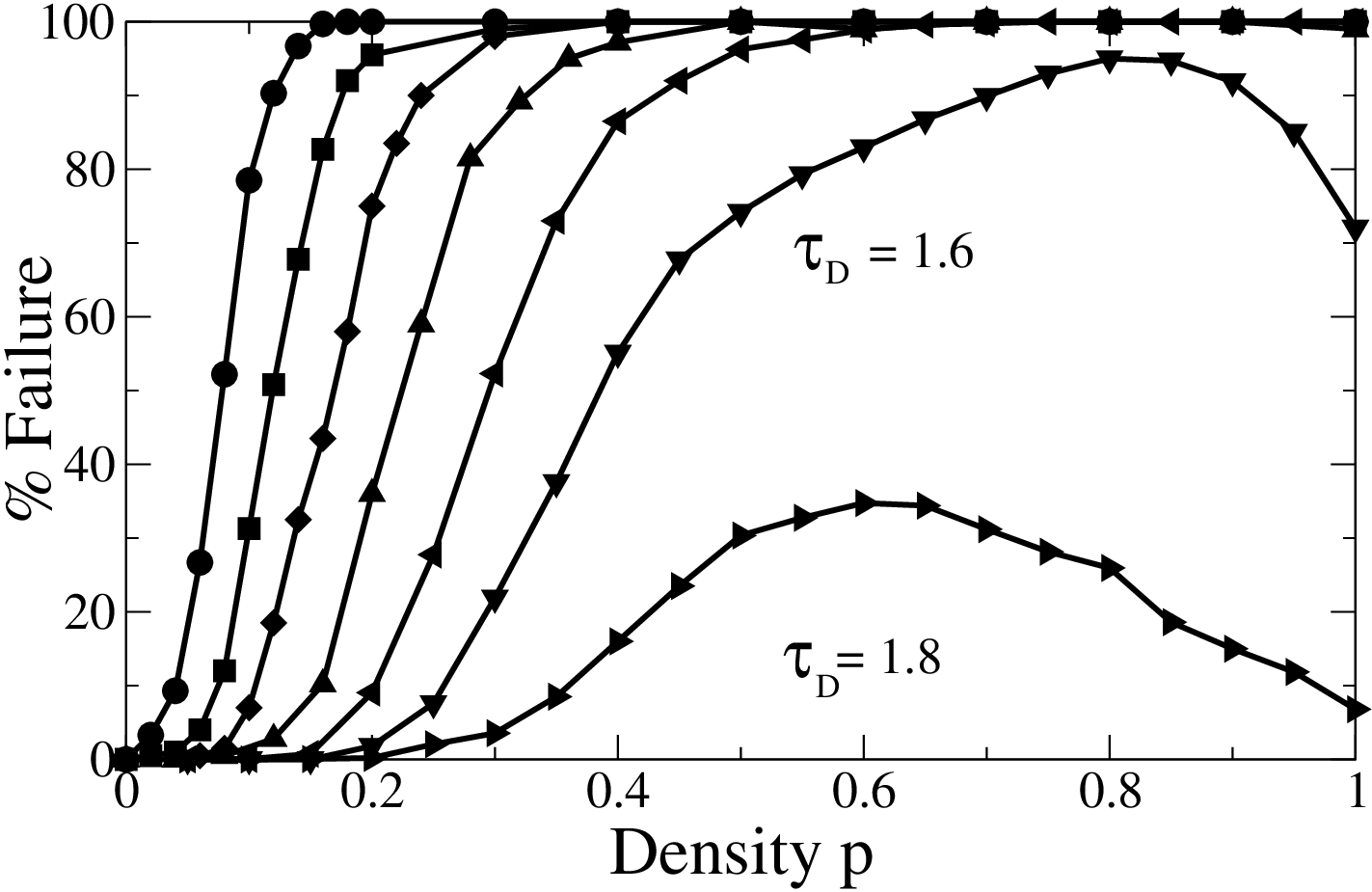}}}
\caption{Failure rates after $t_{max}=2000$  for 
$\tau_{D}=0.6$, 0.8, 1.0, 1.2, 1.4, 1.6, 1.8 (left to right).} \label{fig3a}
\end{figure}

\begin{figure}
\centerline{
\epsfxsize=3.0in{\epsfbox{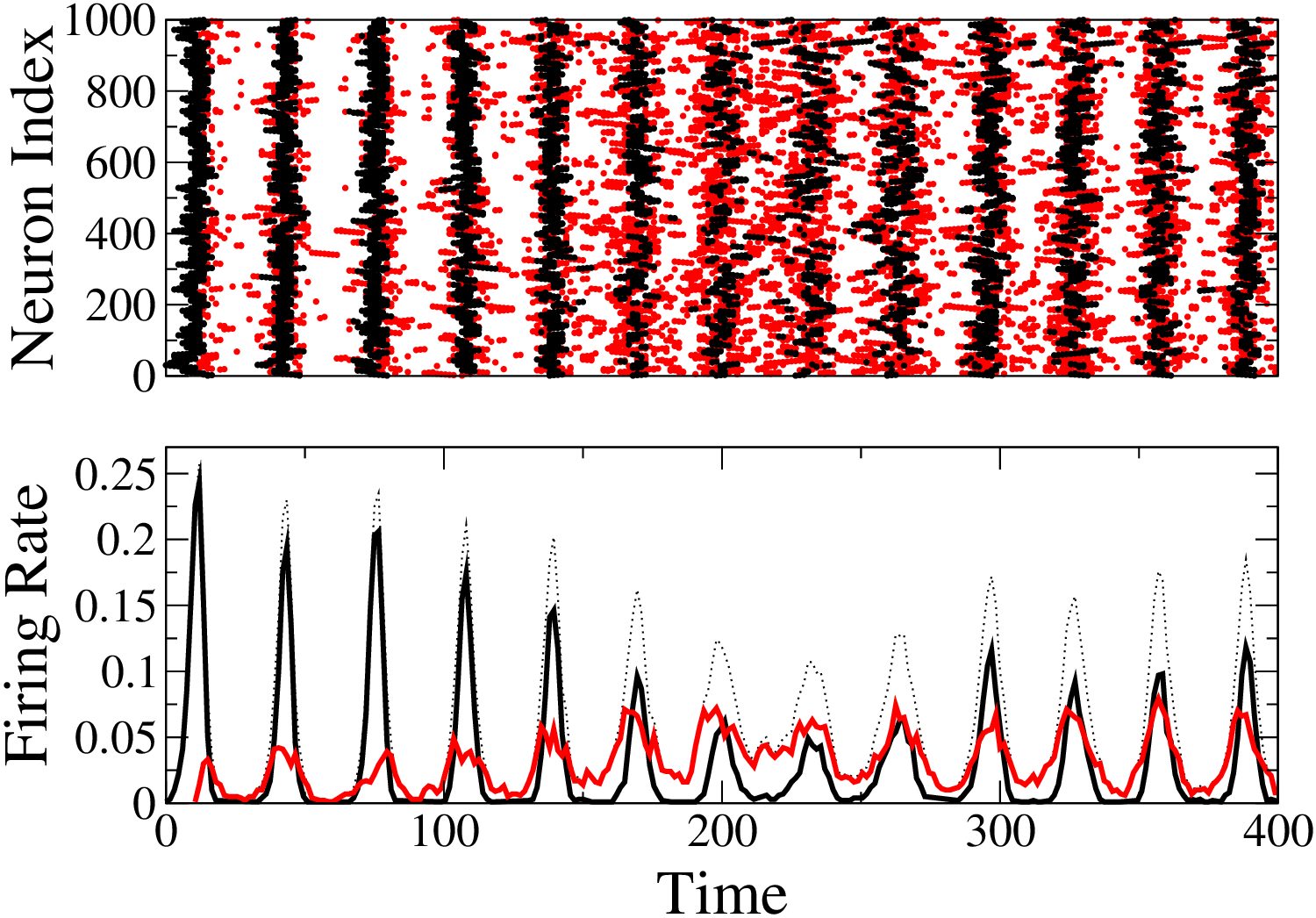}}}
\caption{Raster plot (top) and instantaneous firing rate (bottom)  
of neurons with $ISI > T_R^{(1)}$ (black) and $ISI < T_R^{(1)}$
(red) for $\tau_{D}=1.5$ and $p=1.0$. 
Dotted line is the total firing rate.} \label{fig4}
\end{figure}

\begin{figure}
\centerline{
\epsfxsize=3.0in{\epsfbox{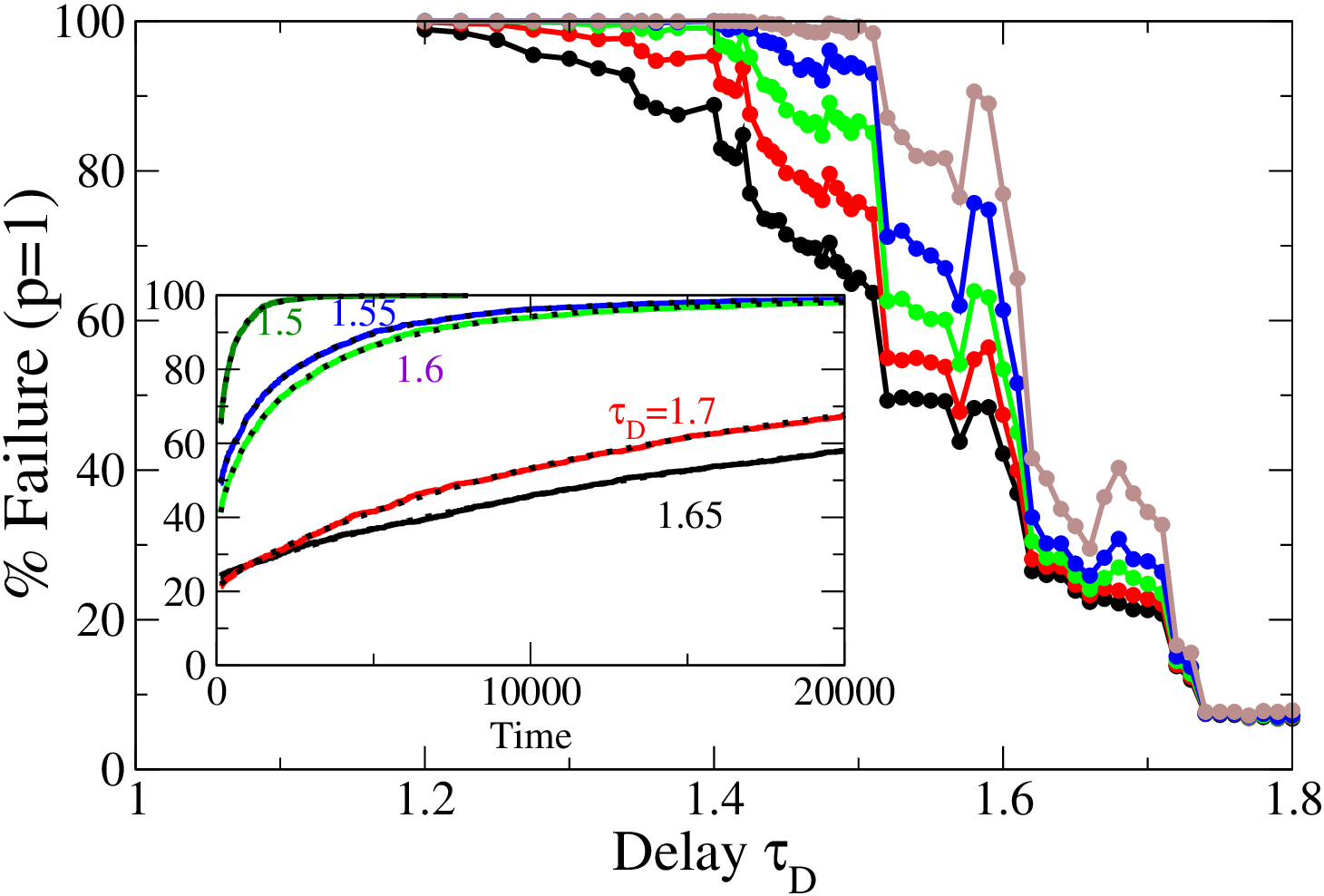}}}
\caption{Failure-time distribution. 
Inset: Cumulative distribution of failure times at $p=1$ for
$1.5 \leq \tau_D \leq 1.7$. 
Main: Failure rates at 5, 10, 20, 40 and 100 multiples of
$T_{R}^{(1)}=28.3$ (black to brown).
Symbols are averages over 2000 realizations.} \label{cyclefail1}
\end{figure}

This well-defined transition to failure occurs only for sufficiently
fast waves, i.e. for short delay $\tau_D$. For larger $\tau_D$
the dynamics of the system become quite more complex, and the
fraction of realizations that fail before a fixed time
($t_{max}=2000$ in fig.\ref{fig3a}) becomes a nonmonotonic 
function of $p$. While at low $p$ the firing patterns are highly
regular (cf. fig.\ref{fig1}) and all failures occur within
one or two cycles of the initial activity, for higher $p$  the
patterns are more disordered (cf. fig.\ref{fig4}) and the
activity can persist for a very long time before failure. Consequently, 
the distribution of failure times, shown in fig.\ref{cyclefail1}, exhibits 
an increasingly long tail for longer delay times.

To understand the persistence of activity beyond $p_{cr}$,
it is important to recognize that the result for the critical
density $p_{cr}(N)$ hinges on the assumption
that each neuron receives but a single excitatory input during
each cycle of network activity. Its recovery time is therefore
given by $T_R^{(1)}$, which sets a lower bound for the interspike
interval (ISI). While this assumption is well satisfied for
small $p$, it does not hold for $p=O(1)$. In fact, the
likelihood that a neuron has $n$ incoming shortcuts follows a
multinomial distribution such that the fraction of neurons receiving
two incoming connections grows from about 0.05\% at $p=0.1$ to
about 30\% at $p=1$. Neurons that receive $n$  
inputs can have a recovery time as short as $T_R^{(n)}$.
Fig.\ref{fig4} reveals that such neurons, with ISIs lower than
$T_R^{(1)}$, play a crucial role in maintaining network activity
in the regime $p\approx 1$. While neurons with ISIs greater than
$T_R^{(1)}$  (marked black in fig.\ref{fig4}) can go through
silent epochs with near-zero activity, neurons with shorter ISIs
(marked red in fig.\ref{fig4}) may fire two or three times during 
a network cycle and carry over the network activity
across these silent epochs.

The persistence of activity beyond $p_{cr}$ depends 
sensitively on $\tau_D$; this reflects the fact that
persistence at higher densities $p$ is due to chains 
or trees of neurons that bridge the silent epochs due to their 
multiple inputs. Since activity
propagates with a fixed speed, a chain of multiple-input neurons
of given length can bridge a time interval proportional
to $\tau_D$. Thus, as $\tau_D$ increases, ever shorter chains can
contribute to bridging a silent epoch of given duration; the
likelihood for failure will decrease accordingly. 
This picture is of course overly simplistic: whether a
\textit{topological} chain can be utilized as a
\textit{dynamical} bridge over a given silent epoch depends on
the amount and timing of the inputs it receives, which in turn
depend on the recent history of the entire network.  
Simulations reveal that the identity of the neurons that form 
the `bridging' dynamical chains varies from cycle to cycle in an
irregular way. This implies that the effective utilization of a
dynamical chain over one cycle does not guarantee its
availability on the next cycle.  Therefore, even systems that
persist for long times may still have a finite probability of
failing. 

The cumulative failure distribution function $F(t)$, shown in
fig.\ref{cyclefail1} for various values of $\tau_D$, exhibits a
long tail and is well fit with stretched exponentials: 
$F(t)=f_{\infty}(\tau_D)-C e^{-\alpha t^{\beta}}$ with $\beta
\approx 0.4$ (dotted lines). Even though the fits are based on
runs up to $t=300,000$ for $\tau_D=1.65$ and $\tau_D=1.7$, they
do not provide a value of $f_\infty (\tau_D)$ accurate enough 
for establishing whether true persistent activity exists for 
a small fraction of the network realizations 
($0.97 \le f_\infty(\tau_D=1.65) \le 1$). Strikingly,
the dependence of $F(t)$ on the delay time exhibits a high
degree of structure, suggestive of `resonances' at values of
$\tau_D$ for which some chains and trees can be optimally
utilized.

\begin{figure}
\centerline{
\epsfxsize=3.0in{\epsfbox{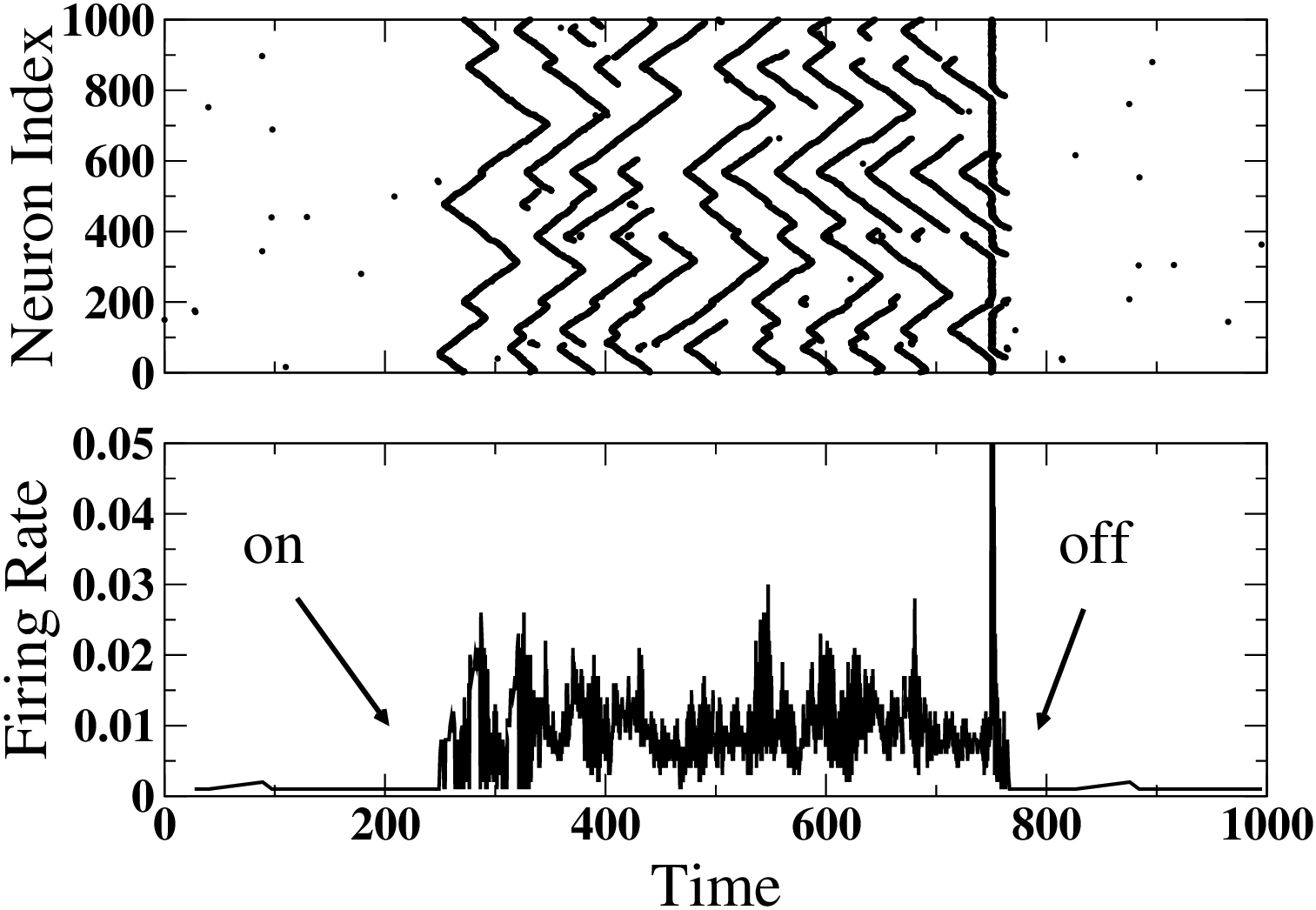}}}
\caption{Raster plot (top) and instantaneous firing rate (bottom)
for $p=0.10$ and $k=5$. About 10 adjacent neurons are stimulated 
synchronously at $t=250$; 
about $20\%$ of the neurons are
activated at $t=750$.} \label{fig5}
\end{figure}

One of the salient features of the emergent dynamics of the model
is persistent
\textit{self-sustained} activity at low densities $p$ of shortcuts. 
In this regime, the network
is in fact bistable between `off' and `on' states, and can be
switched between them with sufficiently large stimuli, as
illustrated in fig.\ref{fig5}. The synchronous stimulation of a 
sufficiently large number of neurons while the network is in the `on' 
state increases the level of activity and effectively
pushes the 
network to the right of the failure transition (cf. fig.\ref{fig2}),
causing a transition into the `off' state. 

To address the fact that real neuronal systems are 
noisy, the simulation shown in fig.\ref{fig5}
includes Gaussian fluctuations in the membrane potential; their
amplitude is chosen so as to cause the neurons to spike
irregularly at a low rate. To keep the noisy spiking of a single
neuron from generating a traveling pulse and initiating the `on'
state, we adjust the synaptic conductance so that several
adjacent neurons must fire in rapid succession in order to
propagate a pulse of excitation. The network topology is modified
accordingly: we extend the local coupling to include up to $2k$
neighbors ($w_{i,i\pm j}=1$ for  $j=1\dots k$) and model
long-range connections via a population of intermediate
excitatory neurons which both receive input from and project to
multiple adjacent neurons. Under these conditions, spontaneous  activity
is highly unlikely to initiate traveling pulses. However, a sufficiently 
large stimulus, synchronous across several neurons, can again  
turn the state of elevated activity on and off (cf. fig.\ref{fig5});
bistability is thus robust with respect to noise.

Network bistability has been hypothesized to be the neural correlate
underlying the type of short-term memory known as working memory in the
prefrontal cortex of monkeys and humans.  Much more realistic and
physiologically plausible models of cortical layers have been studied
within the context of working memory (e.g.
\cite{CoBr00}).  
Yet, not much attention has been given to heterogeneities in network
topology or to long-range  excitatory connections. The work
presented here suggests that closer attention be given to the role 
of connectivity as an additional factor that contributes to the 
generation of the persistent, active state 
associated with working memory. 

In conclusion, we have investigated the effect of incorporating
random unidirectional shortcuts to a one-dimensional network of
locally coupled integrate-and-fire neurons. We find that even a
very low density of shortcuts suffices to generate persistent
activity from a local stimulus through the reinjection of activity
into previously excited domains. As the density of shortcuts is
increased, the substantial decrease in the effective system
size characteristic of small-world networks causes a crossover
into a regime  characterized by failure to sustain activity for
essentially all network configurations. For sufficiently slow
propagation velocities of the activity and sufficiently high
shortcut densities, an intriguing second crossover occurs into a
regime in which the activity still fails but only after often
exceedingly long and strongly chaotic transients. 

The complex dynamical phenomena we find in this extremely simple
model are based on a robust mechanism: propagating pulses of
activity that are sustained by branching and reinjection. We
therefore expect that more realistic models of neuronal
networks, which may include multiple ion channels and continuous
synaptic currents as well as inhibitory coupling, will show
qualitatively the same behavior upon the addition of shortcuts
if they originally support propagating pulses of activity
\cite{GoEr01} that annihilate upon collision.  Preliminary
computations show that the failure transition persists if
shortcuts are not allowed to exceed a maximal length $L_m$, as
long as $L_m \tau_D > T_R^{(1)}$ \cite{RoRiun}. The phenomena
should also be accessible in excitable chemical systems
\cite{MaLi02}, in which the shortcuts could be implemented using
video feedback. 

We gratefully acknowledge support from the Department of Energy
(DE-FG02-92ER14303), NSF grant DMS-9804673, and the NSF-IGERT program
{\it Dynamics of Complex Systems in Science and Engineering}
(DGE-9987577).

\bibliography{journal}
 
\end{document}